\documentclass[twocolumn,aps,superscriptaddress,amsmath,amssymb,preprintnumbers,floatfix]{revtex4-1}

\usepackage[utf8]{inputenc}
\usepackage{amsmath,amssymb}
\usepackage{mathtools}
\usepackage{bm}
\usepackage{times}
\usepackage{graphicx,xcolor,soul,sidecap}
\usepackage[colorlinks,bookmarks=false,citecolor=blue,linkcolor=black,urlcolor=blue,hyperfootnotes=true]{hyperref}
\usepackage[capitalize]{cleveref}
\makeatother

\begin{document}

\title{Multiple Andreev reflections in two-dimensional Josephson junctions \\ with broken time-reversal symmetry}
\author{Linde A.B. {Olde Olthof}}
\thanks{These authors contributed equally}
\affiliation{MESA+ Institute for Nanotechnology, University of Twente, The Netherlands}
\affiliation{Department of Materials Science \& Metallurgy, University of Cambridge, United Kingdom}
\author{Stijn R. {de Wit}}
\thanks{These authors contributed equally}
\affiliation{MESA+ Institute for Nanotechnology, University of Twente, The Netherlands}
\author{Shu-Ichiro Suzuki}
\affiliation{MESA+ Institute for Nanotechnology, University of Twente, The Netherlands}
\affiliation{Department of Applied Physics, Nagoya University, 
Nagoya 464-8603, Japan}
\author{Inanc Adagideli}
\affiliation{MESA+ Institute for Nanotechnology, University of Twente, The Netherlands}
\affiliation{Faculty of Engineering and Natural Sciences, Sabanci University, Istanbul, Turkey}
\affiliation{T\"UB\.ITAK Research Institute for Fundamental Sciences, 41470 Gebze, Turkey}
\author{Jason W.A. Robinson}
\affiliation{Department of Materials Science \& Metallurgy, University of Cambridge, United Kingdom}
\author{Alexander Brinkman}
\email{a.brinkman@utwente.nl}
\affiliation{MESA+ Institute for Nanotechnology, University of Twente, The Netherlands}
\date{\today}

\begin{abstract}
Andreev bound states (ABS) occur in Josephson junctions when the total phase of the Andreev and normal reflections is a multiple of $2\pi$. In ballistic junctions with an applied voltage bias, a quasi-particle undergoes multiple Andreev reflections before entering the leads, resulting in peaks in the current-voltage $I(V)$ curve.
Here we present a general model for Josephson junctions with spin-active interlayers i.e., magnetic or topological materials with broken time-reversal symmetry. We investigate how ABS change the peak positions and shape of $I(V)$, which becomes asymmetric for a single incident angle. We show how the angle-resolved $I(V)$ curve becomes a spectroscopic tool for the chirality and degeneracy of ABS.
\end{abstract}

\maketitle

Andreev reflection is the conversion of an electron into a hole with opposite spin upon reflecting from a superconductor interface \cite{Andreev}. Andreev bound states (ABS) arise when a combination of a number of Andreev and normal reflections fulfills the Bohr-Sommerfeld quantization condition in which the total phase adds up to multiples of 2$\pi$. Renowned examples include ABS that carry the supercurrent between two superconducting leads across a normal metal \cite{Kulik}, the Yu-Shiba-Rusinov bound states that involve scattering from a magnetic impurity \cite{Yu,Shiba,Rusinov}, and the Caroli-de Gennes-Matricon bound state in the core of an Abrikosov vortex \cite{Caroli}. Generally, the required phase quantization is either fulfilled by incorporating spin-active scattering with different phases for the reflection of different spins \cite{Eschrig,Beckmann,SHMS}, or by picking up a phase difference due to an anisotropic order parameter in the superconductor (an unconventional superconductor) \cite{Hu,KashiwayaTanaka,KashiwayaTanaka2}.

At the surface or interface of an unconventional superconductor, surface ABS at zero energy (relative to the Fermi energy) arise when the phase difference between the energy-dependent Andreev reflection of the electron and hole is $\pi$. This has been measured at the surface of a 45$^\circ$ grain boundary junction involving a $d_{x^2-y^2}$ cuprate superconductor \cite{KashiwayaTanaka2} and predicted for the surface of a chiral $p$-wave superconductor \cite{ReadGreen}. Surface ABS become (chiral) Majorana bound states upon lifting the spin degeneracy by breaking time-reversal symmetry in a topological superconductor \cite{Beenakker}, either by a vortex \cite{FuKane2008} or an external magnetic field \cite{Mourik}. The distinguishing feature of a subgap ABS is the zero-bias conductance peak in the tunneling conductance \cite{Hu,Tanaka2005}, which can even become quantized in the Majorana case \cite{Law,Beenakker}.

Here, we theoretically study the influence of (chiral) interface ABS on the current-voltage characteristics of Josephson junctions. Besides a zero-voltage supercurrent, Josephson junctions are characterized by a subgap structure in the finite-bias conductance that arises from multiple Andreev reflections (MAR), usually providing peaks at $2\Delta/n$, where $\Delta$ is the superconducting energy gap, and $n$ is the integer number of times that the electrons or holes traverse the junction before entering the leads \cite{AverinBardas}.  
If a topological insulator (TI) interlayer featuring a magnetic (MTI) barrier is used instead of a normal metal barrier, the subgap state opens an extra conduction channel and the peaks in a one-dimensional (1D) S/TI/MTI/TI/S junction are located at $\Delta/n$ \cite{Badiane2011}. 

\begin{figure}
    \centering
    \includegraphics[width=\linewidth]{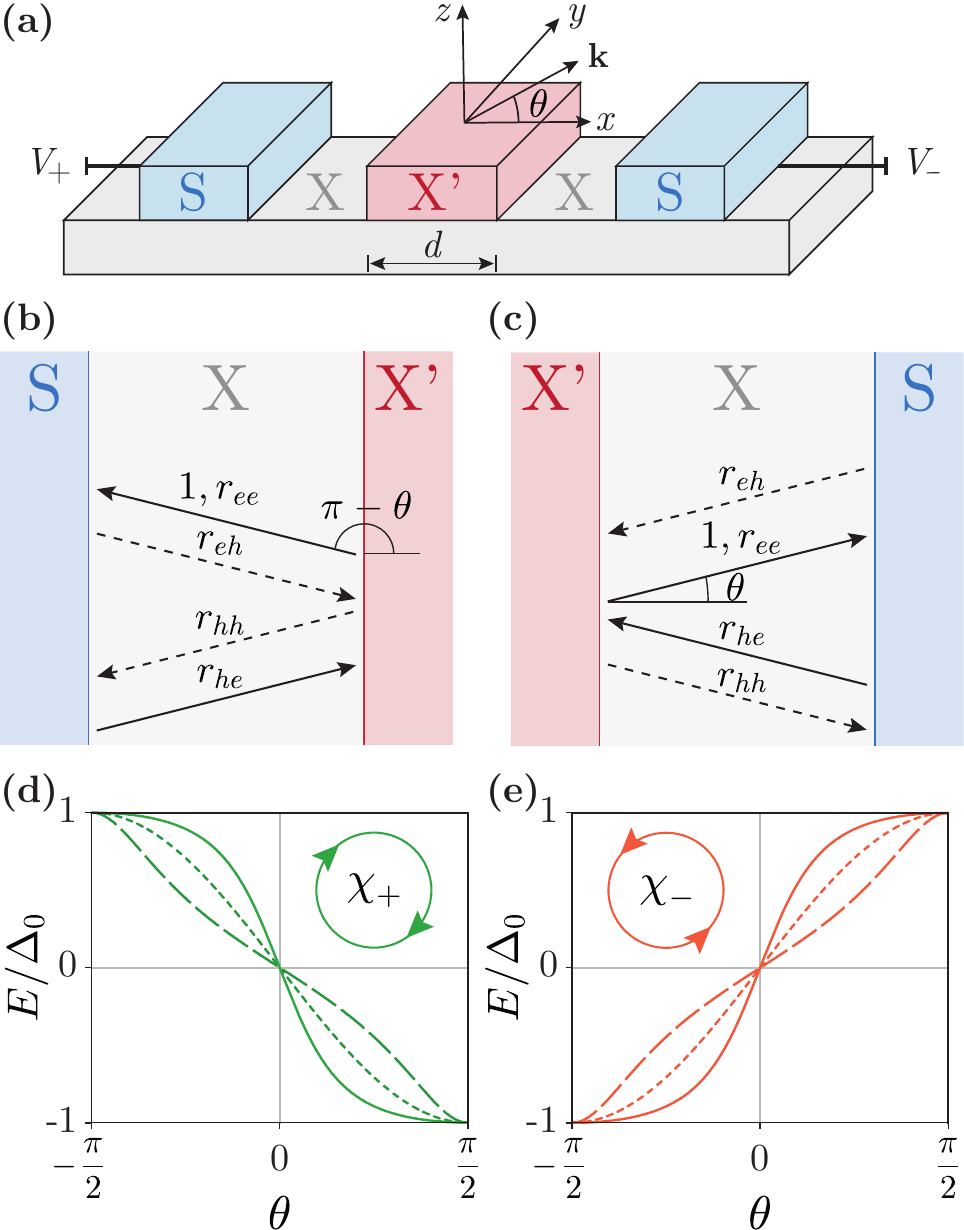}
    \caption{{\bf (a)} Schematic illustration of the S/X/S junction setup, where S is a superconductor and X is a spin-active layer, e.g. a ferromagnetic insulator or a (magnetic) topological insulator. The junction is modelled as S/X/X'/X/S where Andreev reflection occurs in the two X regions which are coupled via a scattering region X' of width $d$. {\bf (b)-(c)} The top view of the S/X/X'/X/S junction split in two half-junctions S/X/X' and X'/X/S, where normal ($r_{ee}$, $r_{hh}$) and Andreev ($r_{eh}$, $r_{he}$) reflection occurs. $\tan\theta = k_y/k_x$ for $k_x,k_y$ being the components of the plane wave momenta.
    {\bf (d)-(e)} In the case of a topological insulator junction (X=TI) with a magnetic tunnel barrier (X'=MTI), the two half-junctions host chiral Majorana modes of opposite chirality $\chi_\pm$. Their bound state levels $E$ as a function of the incident angle $\theta$ are shown for a {\bf (d)} MTI/TI/S and a {\bf (e)} S/TI/MTI junction. The dashed, dotted, and solid lines correspond to $\mu_\text{TI}/m_z = 0.5, 1, 2$, respectively. The other parameters are $m_z = 300\Delta_0$, $\mu_\text{MTI}=0$, $\mu_\text{TI}=\mu_\text{S}$.}
    \label{Fig1}
\end{figure}

We present a generalized model for two-dimensional (2D) Josephson junctions consisting of $s$-wave superconductors (S) and spin-active layers (X) e.g., a TI, a magnetic TI (MTI) or ferromagnetic insulator, see Fig.~\ref{Fig1}a. We structure our quantitative calculations around a 2D S/TI/MTI/TI/S junction of which the S/N/S and 1D S/TI/MTI/TI/S systems are the limiting cases -- after which we generalize our methods to non-topological junctions.
Our findings show a direct link between broken time-reversal symmetry, the presence of ABS and asymmetric angle-resolved $I(V)$ curves.

The approach is as follows: we investigate the existence and energy dependence of ABS in S/X/X' and X'/X/S half-junctions, where X' contains a potential difference or magnetization. We then couple two half-junctions together into a full S/X/S junction and calculate the $I(V)$ spectrum. We show that the features in $I(V)$ are directly linked to the ABS, and discuss how angle-resolved MAR can become a spectroscopic tool for the chiral nature and degeneracy of ABS.
Throughout this paper, the interface normal is along the $x$-axis and we use periodic boundary conditions along $y$ (see Fig.~\ref{Fig1}a). We assume the junction length to be smaller than the coherence length and the elastic mean free path, making the transport coherent and ballistic.

Quasi-particles undergo normal (Andreev) reflection at the X/X' (S/X) interface.
In the X'/X/S junction in Fig.~\ref{Fig1}c, we consider an incoming electron with angle $\theta$ consecutively undergoing Andreev ($r_{eh}$), normal ($r_{hh}$), Andreev ($r_{he}$), and normal ($r_{ee}$) reflection. Fig.~\ref{Fig1}b shows the equivalent process in the other half-junction.
To generalize the reflection processes and incorporate phase differences due to topology and/or magnetism, we introduce a so-called reflection asymmetry phase $e^{i\chi} \coloneqq r_{hh}/r_{ee}^*$ as the ratio between the hole-hole and electron-electron reflection coefficients of X'. 
We compute $r_{hh}$ and $r_{ee}$, by imposing the continuity of the wave function across the junction. The spinor part of the wave function is derived from the Bogoliubov-de Gennes Hamiltonian in the basis $(u_{\uparrow}, u_{\downarrow}, v_{\downarrow},  -v_{\uparrow})^T$,
\begin{equation}
    \hat{H}  = \begin{bmatrix} \hat{h}(\mathbf{k}) & \hat{\Delta} \\ \hat{\Delta}^* & -\sigma_y \hat{h}^*(-\mathbf{k})\sigma_y \end{bmatrix},
    \label{Hamiltonian}
\end{equation}
where $\hat{h}(\mathbf{k}) = f(\mathbf{k}) + \sigma_z m_z - \mu_j \sigma_z $ is the single-particle Hamiltonian. The dispersion, $f(\mathbf{k})$, can include a kinetic term or spin-orbit physics. $\mu_j$ with $j=$ S, X, X' sets the chemical potential in the three regions, and $m_z$ sets the induced magnetic gap. $\hat{\Delta} = \sigma_0 \Delta_0$, with $\Delta_0\in\mathbb{R}$ is the $s$-wave superconducting gap. $\Delta_0$ and $m_z$ are only nonzero in their respective regions. The matrices $\sigma_i$ for $i=0,x,y,z$ are the Pauli matrices.\\

The Hamiltonian~(\ref{Hamiltonian}) obeys particle-hole symmetry. In the absence of a magnetic barrier ($m_z=0$), it is also time-reversal symmetric. This means that the system is placed in symmetry class~BDI when both particle-hole and time-reversal symmetry are present, while it is in class~D when time-reversal symmetry is broken \cite{Ryu}.
Based on the system's symmetries, the topological invariant $Q$ (the number of symmetry-protected edge states present at the Fermi level) per spatial dimension can be calculated through the reflection block of the scattering matrix \cite{Fulga} for the X/X' interface, $ \hat{r} =  \text{diag}(r_{ee},r_{hh})\equiv \text{diag}(r_{ee},e^{i\chi}r^*_{ee})$. In 2D, the topological invariant is a winding number, given by $Q^\text{2D} = \frac{1}{2\pi i } \int_0^{2\pi} dk \frac{\mathrm{d}}{\mathrm{d}k} \log (\det \hat{r})$, where $\det\hat{r}~=~e^{i\chi}$. Details on the symmetry classes and calculation of $Q^\text{2D}$ are provided in Sec.~S1 of the Supplemental Materials~\cite{supp}.

We compute $Q$ for the case of a topological half-junction (X = TI) with and without time-reversal symmetry. In time-reversal symmetric junctions ($m_z=0$), we find $r^*_{ee} = r_{hh}$, such that $e^{i\chi} = 1$ and $Q^\text{2D}=0$, implying that the system is topologically trivial and there are no edge modes. In the case of broken time-reversal symmetry ($m_z\neq 0$), we obtain $Q^\text{2D}=-1$ ($Q^\text{2D}=+1$) for the S/X/X' (X'/X/S) half junction. A topological invariant of $\pm 1$ means that a topologically protected chiral edge mode is present. Importantly, the sign difference of $Q$ between the two half-junctions indicates opposite chirality (winding direction), as illustrated in the inset of Fig.~\ref{Fig1}d,e. 
The protected chiral edge mode in 2D is the nonzero energy chiral Majorana mode originating from the localized zero-energy Majorana bound state present in the 1D channel at the symmetry point $\theta=0$.

Chiral Majorana modes have been predicted in MTI/S junctions \cite{FuKane2008}, and their bound state energies $E_\text{ABS}(\theta)$ were previously found as poles in the conduction \cite{Tanaka}. We compute $E_\text{ABS}(\theta)$ as the energy when the Bohr-Sommerfeld quantization condition, \(\alpha_{r_{ee}} + \alpha_{r_{eh}} + \alpha_{r_{hh}} + \alpha_{r_{he}} = 2\pi n, n\in\mathbb{Z}\), is satisfied for the reflection coefficients depicted in Fig.~\ref{Fig1}b,c. 
For subgap energies, $|E|<\Delta_0$, 
the quantization condition can be written in terms of $\chi$ as $-2\arccos{(E/\Delta_0)} + \chi = 2\pi n $ (Sec.~S2 of the Supplemental Materials~\cite{supp}).
Since $2\arccos{(E/\Delta_0)}$ is bound between 0 and $2\pi$, the condition is met for a nonzero $\chi$. So, the value of $\chi$ dictates whether ABS exist.  
In time-reversal symmetric systems (for instance, when X' is a Fermi surface mismatch barrier), $\chi=0$ and no ABS forms. 
Whereas in time-reversal symmetry breaking systems, $\chi$ is nonzero. The bound state energies vs incident angle for magnetic S/X/X' and X'/X/S junctions are shown in Figs.~\ref{Fig1}d,e.  
At $\theta=0$ (i.e. the 1D limit), the ABS is located at zero energy and is therefore a Majorana bound state. For nonzero angles, the ABS moves away from zero energy and obtains a chirality. 
We recall that the two half-junctions have opposite chirality ($Q=\pm 1$), which results in the $E_\text{ABS}$ having a different sign for a fixed nonzero value of $\theta$. Crucially, this means that in the coupled S/X/X'/X/S junction, for a fixed $\theta$, there are bound states of opposite energy on the left and right side of the X' barrier. \\

To consider the transport in the S/X/X'/X/S junction, we construct the left and right moving wave functions in the two X regions, as eigenfunctions of Eq.~(\ref{Hamiltonian}). We then couple the S/X/X' and X'/X/S junctions via a scattering region X', which is governed by the scattering matrices for electrons and holes
\begin{align}
    &S_e = \begin{bmatrix} r & t \\ t & -\dfrac{r^*t}{t^*} \end{bmatrix},
   && S_h = \begin{bmatrix} e^{i\chi} r^* & t^* \\ t^* & -e^{-i\chi}\dfrac{rt^*}{t} \end{bmatrix}, \label{eq:scattermatrix}
\end{align}
where $r\equiv r_{ee}$ and $t\equiv t_{ee}$ are the electron-electron reflection and transmission coefficients for the barrier X' and we used $r_{hh}/r_{ee}^* = e^{i\chi}$. 
Two known limits of the scattering matrices are $e^{i\chi}=1$ for a S/N/S junction \cite{AverinBardas} and $e^{i\chi}=-1$ for a 1D ferromagnetic S/TI/MTI/TI/S junction \cite{Badiane2011}.
We consider a potential difference $eV$ in the X' region, such that in the MAR picture \cite{AverinBardas,Badiane2011}, every time an electron passes from left to right, crossing X', its energy increases by $eV$, while the hole energy increases when it passes in the opposite direction. Consequently, the wave functions are superpositions of states with energy $E+2n eV$ where $E$ is the quasi-particle energy and $n$ the number of Andreev reflections. The Andreev reflection coefficient $a_n$ changes accordingly to $a_n\equiv r_{eh}(E+ neV)$. We note that the choice of basis results in equal Andreev reflection coefficients $a_n$ at the left and right S interface \footnote{In our chosen basis, $a_n$ is equal to the reflection coefficients $r_{eh} = r_{he} = (E-\sqrt{E^2 - \Delta^2_0})/(\Delta_0)$, see Sec.~S2 of the Supplemental Materials~\cite{supp}.}, which is crucial for the MAR calculations. We derive the scattering equations, generalize the MAR recurrence relations \cite{AverinBardas} and compute $I(V)$ based on the amplitudes of the superimposed wave functions. This approach is the key technical finding in this work; details are provided in Sec.~S3 of the Supplemental Materials~\cite{supp}.\\

\begin{figure}
    \centering
    \includegraphics[width=\linewidth]{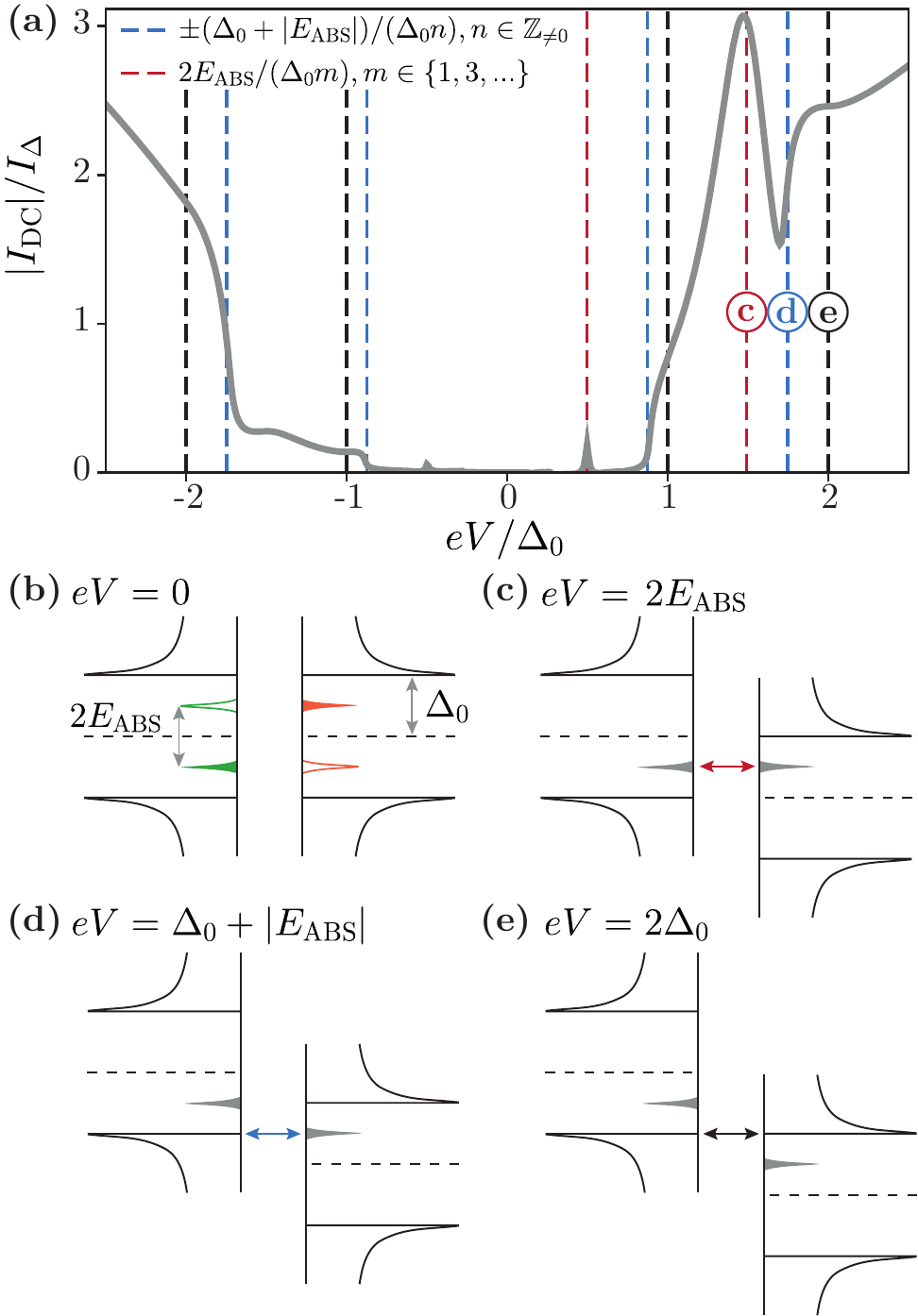}
    \caption{{\bf (a)} An asymmetric $I(V)$ curve of a S/TI/MTI/TI/S junction for a single incident angle $\theta = 0.45 \pi$, with corresponding $E_{\text{ABS}}/\Delta_0 = 0.75$. $I_{\mathrm{DC}}$ is normalized by $I_\Delta = De\Delta_0/h$, for a transparency $D = 0.1$. The other parameters are $\mu_{\text{TI}}/m_z\sim 0.7$ with $m_z/\Delta_0=300$, $\mu_\text{S}=\mu_\text{TI}$, $\mu_\text{MTI}=0$, and the MTI barrier width is $d= 1.5\hbar v_F/m_z$.
    {\bf (b)} The density of states of the two superconductors including the subgap ABS. The filled (empty) ABS positions correspond to a positive (negative) incident angle. The bias voltage $eV$ shifts the density of states of the right superconductor relatively downwards. When considering only the filled ABS for $\theta>0$, transport occurs with the alignment of {\bf (c)} ABS-ABS, {\bf (d)} ABS-continuum and {\bf (e)} continuum-continuum.}
    \label{Fig3}
\end{figure}

\begin{figure*}
    \centering
    \includegraphics[width=.9\linewidth]{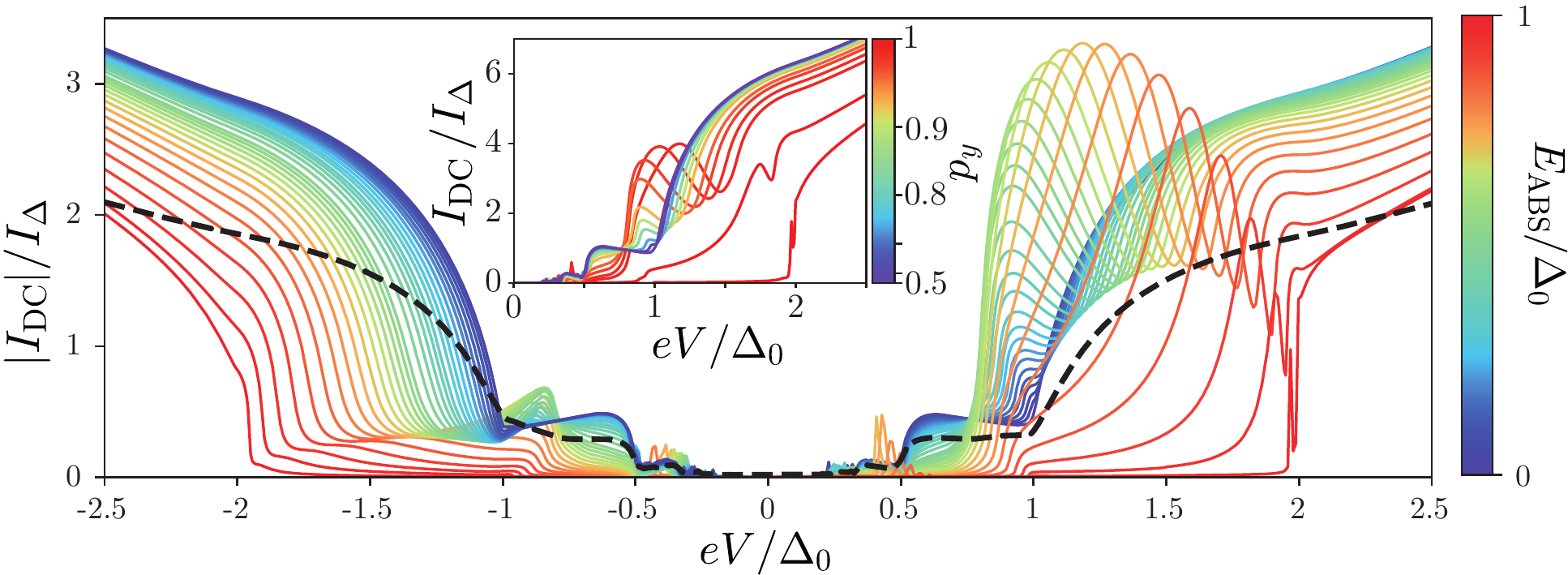}
    \caption{
    The angle-resolved asymmetric $I(V)$ curves for a S/TI/MTI/TI/S junction normalized by $I_\Delta = De\Delta_0/h$. Each curve corresponds to a single incident angle $\theta\in (0,\pi/2)$, with corresponding bound state energy $E_\text{ABS}(\theta)$. The black dashed line is the angle average obtained in the lateral 2D junction limit. The parameters are $\mu_{\text{TI}}/m_z\sim 0.7$, with $m_z/\Delta_0=300$,  $\mu_\text{S}=\mu_\text{TI}$, $\mu_\text{MTI}=0$, the transparency ranges from $D=0.005-0.2$, and the MTI barrier width is $d= 1.5\hbar v_F/m_z$.
    \textbf{Inset:} Nanowire limit. Each $I(V)$ curve corresponds to a single  (normalized) quantized $p_y=\sin\theta$ channel where the current is estimated by $I (-\theta) + I(\theta)$. The graph is identical for $\pm eV/\Delta_0$.
    }
    \label{Fig4}
\end{figure*}

First, we investigate the angle-resolved MAR spectra for a 2D S/TI/MTI/TI/S Josephson junction (Fig.~\ref{Fig3}). In a trivial junction (e.g. S/N/S), there are no states inside the gap, electrons (holes) undergo MAR until they have gained enough energy to leave the gap at $eV= +(-)2\Delta_0$ \cite{AverinBardas}. The presence of an ABS in the gap gives rise to extra conduction channels \cite{Badiane2011}.
When the ABS aligns with the ABS on the other side ($eV=2E_\text{ABS}$) \cite{notesign} or the continuum ($eV=\Delta_0+|E_\text{ABS}|$), additional features appear in the $I(V)$ curve.
Furthermore, due to the nature of MAR, higher-order features appear for successive Andreev reflections. Generally, features in the $I(V)$ curve are expected at $eV=(\Delta_0+|E_\text{ABS}|)/n$, with $n\in\mathbb{Z}$, stemming from alignment of the ABS with the continuum; and at $eV=2E_\text{ABS}/m$, for odd $m\in\mathbb{N}$, stemming from alignment of the two ABSs. These alignments are illustrated in Fig.~\ref{Fig3}. 
We note that the modes with opposite chirality on the left and right side of the MTI barrier are retained in the coupled Josephson junction (Sec.~S1 of the Supplemental Materials~\cite{supp}). The energy asymmetry resulting from these modes of opposite chirality dictates that $m$ must be odd. This can be seen by considering an electron initially incoming from the left subgap state. For it to scatter to the empty subgap state on the other side of the barrier it can only traverse the system an odd number of times, gaining an odd multiple of $eV$ in energy.  

The asymmetry of the bound state energies due to the opposite chirality for the left and right half-junctions in Fig.~\ref{Fig1}d,e gives rise to the asymmetric $I(V)$ curve in Fig.~\ref{Fig3}. For a fixed $\theta$, a positive bias voltage $eV$ aligns different levels (associated with higher order MAR resonances) than a negative bias. In the latter case, the density of states in Fig.~\ref{Fig3}c-e shift in the opposite direction, the $eV = 2E_{\textup{ABS}}$ levels never align and the associated features in $I(V)$ are absent for $eV<0$. The $I(V)$ curve for $-\theta$ is the vertical mirror image of Fig.~\ref{Fig3}.

The value $e^{i\chi}$ in the scattering matrices~(\ref{eq:scattermatrix}) indicates whether ABS are present; $e^{i\chi}=1$ means there are no ABS and results in a trivial $I(V)$ curve (meaning no additional subgap features), whereas $e^{i\chi}= -1$ indicates a non-trivial $I(V)$ curve with asymmetric peaks, as shown in Fig.~\ref{Fig3}a. By changing $\theta$, we smoothly transition from the trivial to non-trivial regime \footnote{The value at which the transition happens depends on the barrier strength $\mu_\text{TI}/m_z$. In the strong barrier limit (low $\mu_\text{TI}/m_z$), the system resembles two mostly isolated half-junctions with a large non-trivial regime, whereas for a weak barrier the non-trivial regime is confined to $\theta=0$.}.

Fig.~\ref{Fig4} shows the $I(V)$ spectra of a S/TI/MTI/TI/S junction for positive $\theta$ ranging from $0$ to $\tfrac{\pi}{2}$, with corresponding positive $E_\text{ABS}$ \cite{notesign}. Two limiting cases are the red curve with the ABS near the continuum ($E_\text{ABS}= \Delta_0$ and $e^{i\chi}= 1$) for which we observe the strong resonance step near $2\Delta_0$ as in the trivial S/N/S case \cite{AverinBardas}; and the navy curve describing ABS positions in the middle of the gap ($E_\text{ABS}= 0$ and $e^{i\chi}=-1$) for which we obtain the topological 1D S/TI/S limit \cite{Badiane2011}.
The intermediate curves look strikingly different. For negative voltages, there is a gradual transition from one limit to the other, whereas the positive $eV$-side features the distinct additional $2E_{\text{ABS}}/m$ peaks for odd $m$ due to the asymmetric ABS. Again, for negative incident angles, we obtain the vertical mirror image of Fig.~\ref{Fig4}. 

We now consider the consequences of the presence of ABS and their effect on the $I(V)$ spectra in realistic experimental setups, e.g. lateral junctions and nanowires.
Specialized setups for measuring particular Andreev-reflection angles \cite{Xiao,Mortensen} could reveal asymmetry in $I(V)$. In 2D samples (lateral junctions on thin films), however, one generally measures the angle-averaged $I(V)$ and the asymmetric features disappear (see the black dashed line in Fig.~\ref{Fig4}). 
Throughout this work, we assumed an infinite junction in the $y$-direction, meaning that there is a continuum of $p_y$ channels and every incident angle $\theta$ is allowed. To apply the MAR scheme to nanowires \cite{Mourik,Oreg,Deng}, we instead consider a cylindrical geometry, where, due to the confinement, we obtain a set of allowed quantized $p_y$ values. Per confined $ p_y$ value, only the corresponding $\theta$ and $-\theta$ channels are present, and we estimate the current through the nanowire by $\sim I(\theta)+I(-\theta)$. Contrary to the lateral junction, the subgap resonances at $eV = 2E_{\mathrm{ABS}}$ are retained in the nanowire (see the inset of Fig.~\ref{Fig4}).

The proposed MAR scheme generalizes to non-topological Josephson junctions. Subgap states are present in any $s$-wave Josephson junction with broken time-reversal symmetry, but the (a)symmetric nature of the ABS is not universal.
In topologically trivial systems the ABS are degenerate (on both sides of the barrier), and thus no energy asymmetry is present in the junction. The corresponding angle-resolved $I(V)$ curves are therefore symmetric, as seen in, e.g. ferromagnetic Josephson junctions \cite{Eschrig,Beckmann,SHMS}. 
In topological systems, a single mode of a separated pair of chiral Majorana modes is confined topologically to either the top or bottom surface of the TI. Since the considered junction is located on the top surface (see Fig.~\ref{Fig1}a), a single subgap state is present on each side of the barrier, giving rise to the asymmetric $I(V)$ curves.

The MAR scheme can also be generalized to Josephson junctions with unconventional superconductors, which are characterized by an anisotropic order parameter with a phase -- e.g. $p_x$-wave as in the Kitaev chain \cite{Kitaev} and $d$-wave high-$T_\text{c}$ cuprates \cite{KashiwayaTanaka,KashiwayaTanaka2}. Unconventional superconductors can have a range of exotic properties such as intrinsic chirality -- which ensures the existence of chiral bound states -- or intrinsically broken time-reversal symmetry -- which eliminates the need for a magnetic barrier.
To implement unconventional superconductivity in our model, we recall that the choice of basis is crucial to get equal Andreev reflection coefficients $a_n$ at the left and right S interface. One can construct a unitary transformation to transfer the phase from the order parameter to $e^{i\chi}$ such that the $a_n$ remains equal at both S interfaces and the MAR scheme is still valid, see Sec.~S4 of the Supplemental Materials~\cite{supp} for details.\\

In conclusion, we have investigated the emergence of (chiral) ABS in 2D Josephson junctions with magnetic and/or topological interlayers and studied their effect on calculated $I(V)$ spectra.
Any Josephson junction with broken time-reversal symmetry features ABS in the density of states. When these align with ABS on the other side or the continuum, a conduction channel opens which appears as a peak in the $I(V)$ curve. This directly links the $I(V)$ curve to the ABS energies.

In topological systems, a single topologically protected ABS is present, which obtains a chirality (winding number) in 2D. The S/TI/MTI/TI/S junction features bound states of opposite chirality on either side of the MTI barrier and the corresponding bound state energies are inverted. This energy asymmetry is responsible for the asymmetric $I(V)$ curve.
We have investigated two limits of the S/TI/MTI/TI/S $I(V)$ curve. In lateral 2D junctions where one experimentally obtains an angle-averaged $I(V)$ curve, the asymmetry disappears but non-trivial steps related to the presence of subgap states remain. In the nanowire limit, the distinct peaks, which are an artefact of the present asymmetric chiral Majorana modes, are robust for quantized $p_y$ channels.

The concept of non-reciprocity has regained interest in the field of superconductivity as a potential probe for broken symmetries \cite{Tokura}. In the S/TI/MTI/TI/S case, the non-reciprocity arises from the energy asymmetry as a function of $\theta$ between the two emergent bound states of opposite chirality on either end of the MTI barrier. Particle-hole symmetry is not violated in this case since the energy of the subgap state also inverts as $\theta$ is inverted. We propose angle-resolved ABS spectroscopy to resolve the predicted asymmetry.

L.A.B.O.O. and J.W.A.R. were supported by the EPSRC through the Core-to-Core International Network ``Oxide Superspin''  (EP/P026311/1) and the ``Superconducting Spintronics'' Programme Grant (EP/N017242/1). L.A.B.O.O. also acknowledges support from the Doctoral Training Partnership Grant (EP/N509620/1).
S.-I.~S. is supported by JSPS Postdoctoral Fellowship for Overseas Researchers and a Grant-in-Aid for JSPS Fellows (JSPS KAKENHI Grant No. JP19J02005). We acknowledge useful discussions with Alexander Golubov. 

\let\oldaddcontentsline\addcontentsline
\renewcommand{\addcontentsline}[3]{}
\bibliography{ref}

\begin{thebibliography}{33}
\bibitem [{not()}]{notesign}%
  \BibitemOpen
  \href@noop {} {}\bibinfo {note} {The bound state energy $E_\text{ABS}$ is
  defined in the MTI/TI/S half-junction. In the full S/TI/MTI/TI/S junction
  $E_\text{ABS}$ is positive for $\theta>0$.}\BibitemShut {Stop}%
\end{thebibliography}
\let\addcontentsline\oldaddcontentsline

\end{document}